\documentclass[aps,pre,amsmath,amsfonts,amssymb,nofootinbib,twocolumn]{revtex4}

\usepackage{graphicx}
\usepackage{eucal}
\newcommand{\me}{\mathrm{e}}
\newcommand{\md}[1]{\mathrm{d}#1}

\begin{document}
\title{Quantum-dot Carnot engine at maximum power}

\author{Massimiliano Esposito}
\affiliation{Center for Nonlinear Phenomena and Complex Systems,
Universit\'e Libre de Bruxelles, CP 231, Campus Plaine, B-1050 Brussels, Belgium.}
\author{Ryoichi Kawai}
\affiliation{Department of Physics, University of Alabama at Birmingham, 1300 University Blvd. Birmingham, AL 35294-1170, USA.}
\author{Katja Lindenberg}
\affiliation{Department of Chemistry and Biochemistry and BioCircuits Institute, University of California,
San Diego, La Jolla, CA 92093-0340, USA.}
\author{Christian Van den Broeck}
\affiliation{Hasselt University, B-3590 Diepenbeek, Belgium.}

\date{\today}

\begin{abstract}
We evaluate the efficiency at maximum power of a quantum-dot Carnot heat engine.
The universal value of the coefficients at the linear and quadratic order in the
temperature gradient are reproduced. Curzon-Ahlborn efficiency is recovered in
the limit of weak dissipation. 
\end{abstract}

\maketitle

\section{Introduction}

The purpose of a heat engine is to transform an amount of heat $Q_h$,
extracted from a hot reservoir at temperature $T_h$, into an amount of work $W$.
The efficiency $\eta=W/Q_h$ for doing so is at most equal to Carnot efficiency:
$\eta \le \eta_{c}$, with $\eta_{c}=1-T_c/T_h$. Here $T_c$ is the temperature of
a second, cold reservoir $T_c \le T_h$, in which the remaining energy $Q_h-W$ is
deposited. The equality is reached for reversible operation, implying that the
corresponding power output is zero. Curzon and Ahlborn were amongst the
first to study the question of efficiency at maximum power~\cite{CA}. By
considering a simple modification of the Carnot engine and after applying the
so-called endo-reversible approximation (neglecting dissipation in the auxiliary
system), they found the following beautiful expression for the efficiency at
maximum power: $\eta_\text{\sc
ca}=1-\sqrt{T_c/T_h}=1-\sqrt{1-\eta_{c}}=\eta_{c}/2+\eta_{c}^2/8+ \cdots$.
While this formula appears to describe rather well the efficiency of actual
thermal plants, and is close to the efficiency at maximum power for several
model systems, it is neither an exact nor a universal result, and it is neither
an upper nor a lower bound. It has therefore come as a surprise that a number of
universal predictions can be made about the expansion of the efficiency at
maximum power in terms of $\eta_{c}$.  In the regime of linear response, i.e.,
at first order in $\eta_{c}$, it is found that the efficiency at maximum power
is at most half of the Carnot efficiency~\cite{VdB05}. In other words, the CA
efficiency is an upper bound at the level of linear response.  The proof was
given for systems operating under steady-state conditions. The upper bound is
reached for so-called strongly coupled systems, i.e., systems in which the heat
flux and the work-producing flux are proportional to each other. More recently,
it has been shown that  the quadratic coefficient, equal to $1/8$,  is also
universal for strongly coupled systems in the presence of an additional
left-right symmetry in the system~\cite{EspoPRL09}.  Furthermore, the
universality of the coefficients is a direct consequence of the
time-reversibility of the underlying physical laws. The coefficient $1/2$
derives from the symmetry of the Onsager matrix. The coefficient $1/8$ can be
seen as the implication of Onsager symmetry at the level of nonlinear response.

The above universality predictions have been confirmed in a number of
steady-state model systems involving classical particles~\cite{Tu08},
fermions~\cite{EspoEPL09a}, and bosons~\cite{EspoPRB09}. Universality has also
been observed in variants of the CA model based on Carnot cycles performed in
finite time, even though finding the optimal driving protocol maximizing work
extraction can be notoriously
difficult~\cite{schmiedl07,schmiedl08,then08,gomez-marin08,Okuda09a,Okuda09b,
Okuda08}. The connection with the steady-state analysis has been
clarified by
identifying the Onsager coefficient for a finite-time Carnot cycle in the linear
regime \cite{Okuda09a}. Furthermore, agreement with the universal quadratic
coefficient has also been observed in a Carnot cycle based on a (classical)
Brownian particle in an harmonic trap~\cite{schmiedl08}.

In this paper we provide a full analysis of a thermal engine undergoing a Carnot
cycle, with the auxiliary system consisting of a single-level quantum dot
that is switched between a hot and a cold reservoir. We show that the efficiency
at maximum power is again consistent with the above discussed universality.
Furthermore, CA efficiency at maximum power is obtained exactly at all orders in
$\eta_{c}$ in the limit of low dissipation, which is distinct but similar to the
case considered in the original CA paper.

\section{Model}\label{sec:model}

Our heat engine model consists of a single-level quantum dot interacting with a
metallic lead through a tunneling junction. The quantum dot is assumed to have a
single energy level $\varepsilon$ near the Fermi level of the lead while other
levels in the dot do not contribute to the processes described below. The state
of the system is specified by the occupation probability $p(t)$ of having an
electron in the dot. The lead plays the role of a thermal bath at temperature
$T$ and chemical potential $\mu$. Electrons are assumed to thermalize
instantaneously upon tunnelling into the lead.

When the energy level $\varepsilon$ is modulated by an external agent according
to a given protocol, a certain amount of energy, positive or negative, flows
into the system in the form of work and/or heat. In the case of an occupied
level, an amount of work equal to $(\varepsilon_f-\mu_f)-(\varepsilon_i-\mu_i)$
is delivered to the system, where the subscripts $f$ and $i$ refer to final and
initial values. When the electrons at energy level $\varepsilon$ tunnel in
(out), an amount of heat equal to $Q=\varepsilon-\mu$ ($Q=-\varepsilon+\mu$) is
extracted from the bath.

\begin{figure}
 \includegraphics[width=3.3in]{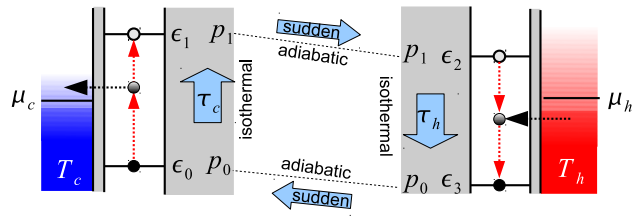}
 \caption{A Carnot cycle of the model heat engine consisting of a single-level quantum dot interacting with a lead through a tunnleing junction.}
 \label{fig:model}
\end{figure}

The basic problem that we address is the finite-time performance of this
engine as it runs through the following four standard stages of a Carnot
cycle (also see figure \ref{fig:model}):

\begin{itemize}
\item [I]  {\it Isothermal process}
\\
The quantum dot is in contact with a cold lead at temperature $T_c$ and chemical
potential $\mu_c$. The energy level is raised from $\varepsilon_0$ to
$\varepsilon_1$ according to a certain protocol during a time interval of
duration $\tau_c$.   Both work and heat are exchanged during this process.

\item [II]  {\it Adiabatic process}
\\
The quantum dot is disconnected from the cold lead, and the quantum level is
shifted from $\varepsilon_1$ to a new level $\varepsilon_2$. Since the quantum
dot is thermodynamically isolated, the population of the level does not change
during this process. Hence, there is no heat exchange. However, the change of
the energy level releases a corresponding amount of work. We assume that the
operation time of this step is very short, in particular negligibly small
compared to that of the isothermal processes.

\item [III] {\it Isothermal process}
\\
The dot is connected to the hot lead at temperature $T_h$ and chemical
potential $\mu_h$. The energy level is lowered from $\varepsilon_2$ to
$\varepsilon_3$ based on another protocol during a time interval of length
$\tau_h$.  Just as in step I,  both heat and work are exchanged.

\item [IV] {\it Adiabatic process}
\\
The system is again disconnected from the lead and the level is restored from $\varepsilon_3$ to the initial level $\varepsilon_0$, at the cost of a corresponding amount of work.  Afterwards, the dot is reconnected to the cold lead. Again, we assume that the operation time of this process is negligibly small.
\end{itemize}

The above procedure defines one cycle of the thermal engine, requiring a total
time $\tau_c+\tau_h$. The protocols in steps I and III must be designed in such
a way that the thermodynamic state of the system, in our case the occupation
probability $p$ of the quantum level, returns to the same initial value
after every cycle. Since there is no change in occupation probability during the
adiabatic stages II and IV, the change in occupation probability from, say $p_0$
to $p_1$, during process I, must necessarily be compensated by a change back
from $p_1$ to $p_0$ during process III.

The time evolution of the occupation probability $p(t)$ for the state of the quantum dot in contact with a lead at temperature $\beta^{-1}$ ($k_b=1$) obeys the following quantum master equation:
\begin{equation}
 \dot{p}(t)= -\omega_a(t) p(t) + \omega_b(t) [1-p(t)],
\label{eq:master}
\end{equation}
where the $\omega_a$ and $\omega_b$ are transition rates. In the wide-band
approximation, these rates are given by
\begin{subequations}
\label{eq:rates}
\begin{eqnarray}
 \omega_a &=& \frac{C}{\me^{-\beta[\varepsilon(t)-\mu(t)]}+1} \label{eq:rate1}\\
 \omega_b &=& \frac{C}{\me^{+\beta[\varepsilon(t)-\mu(t)]}+1} \label{eq:rate2},
\end{eqnarray}
\end{subequations}
where $C$ is a rate constant. Noting that raising the energy level is equivalent to lowering the chemical potential, we introduce an effective energy level $\epsilon \equiv \varepsilon-\mu$. The master equation~(\ref{eq:master}) can now be rewritten as
\begin{equation}\label{eq:master2}
 \dot{p}(t) = -C p(t) + \frac{C}{\me^{\beta \epsilon(t)}+1} .
\end{equation}
The effective level varies along the Carnot cycle as $\epsilon_0=\varepsilon_0-\mu_c \rightarrow \epsilon_1=\varepsilon_1-\mu_c \rightarrow  \epsilon_2=\varepsilon_2-\mu_h \rightarrow \epsilon_3=\varepsilon_3-\mu_h$. Note that the change in the chemical potential is included in the jump of the effective level during processes II and IV.

We next turn to the thermodynamic description of the model. We use the convention that heat entering the system is (like work) positive. The internal energy of the system at time $t$ is
\begin{equation}
\mathcal{E}(t) = \mathcal{U}(t) - \mu \mathcal{N}(t) = \epsilon(t) p(t),
\label{internalenergy}
\end{equation}
where
\begin{equation}
\mathcal{U}(t) = \varepsilon(t) p(t), \quad \mathcal{N}(t) = p(t).
\label{pieces}
\end{equation}
The rate of change of the internal energy,
$\dot{\mathcal{E}}$, is the sum of two parts, namely, a work flux
 $\dot{\mathcal{W}}$ and a heat flux $\dot{\mathcal{Q}}$,
\begin{subequations}
\begin{eqnarray}
\label{workflux}
\dot{\mathcal{W}} &\equiv& \dot{\epsilon}p = \dot{\varepsilon} p -\dot{\mu} p\\
\label{heatflux}
\dot{\mathcal{Q}} &\equiv& \epsilon\dot{p} = \varepsilon \dot{p} - \mu \dot{p}.
\end{eqnarray}
\end{subequations}
Note that the particle exchange contributes to the heat flux [last term in Eq.~(\ref{heatflux})]. When the energy level is below the chemical potential, the direction of heat flow is opposite to the direction of tunnelling.

The net total work and net total heat during the process of duration $\tau$ are obtained as functionals of the occupation probability,
\begin{subequations}
\begin{eqnarray}
\mathcal{W}[p(\cdot)] &=& \int_{0}^{\tau}  \dot{\epsilon}(t) p(t) \md t \label{eq:W} \\
\mathcal{Q}[p(\cdot)] &=& \int_{0}^{\tau}  \epsilon(t) \dot{p}(t) \md t
\label{eq:Q}.
\end{eqnarray}
\end{subequations}
For cyclic processes, we have $\mathcal{E}(0)=\mathcal{E}(\tau)$ and hence $\mathcal{W}+\mathcal{Q}=0$. For mathematical simplicity, we evaluate power using net heat instead of net work:
\begin{equation}
 \mathcal{P}=\frac{-\mathcal{W}}{\tau}=\frac{\mathcal{Q}}{\tau}\, .
\end{equation}

\section{Optimization: General Case}

Our goal is to maximize the power output and to evaluate the corresponding
efficiency. Power is a complicated functional of the time-dependent protocols in
stage I and III, and an exact analytical analysis looks difficult at first
sight. The optimization can however be done in two steps. First, we fix
parameter values, $\tau_c$, $\tau_h$, $p_0$ and $p_1$ and maximize the power
with respect to the functional space of $\epsilon(t)$. Since the total operation
time is fixed, we just need to maximize the heat. Next, we further maximize the
power with respect to the remaining degrees of freedom $\tau_c$, $\tau_h$, $p_0$
and $p_1$. The problem of maximizing heat or minimizing work for a single-level
quantum dot moving between given initial and final energy states has already
been analysed in detail in \cite{EspoEPL09b}. We reproduce the crucial
steps of this analysis for self-consistency.

To find the protocol that maximizes the heat, we do not search directly for the optimal schedule  $\epsilon(t)$, but identify the optimal occupation probability $p(t)$. This is done by  expressing $\epsilon(t)$  in terms of
$p(t)$ and $\dot{p}(t)$, and rewriting the heat, Eq.~(\ref{eq:Q}), as a functional of $p(t)$ and $\dot{p}(t)$:
\begin{equation}
\beta \mathcal{Q}[p(\cdot)]=\int_{0}^{\tau} \mathcal{L}(p,\dot{p}) \md t,
\end{equation}
where
\begin{equation}
 \mathcal{L}\equiv \ln \left [ \frac{1}{p(t)+\dot{p}(t)}-1 \right ] \dot{p}(t).
\end{equation}
The extremum is found via the standard Euler-Lagrange method, leading, after integration, to
\begin{equation}
 \mathcal{L} - \dot{p} \frac{\partial \mathcal{L}}{\partial \dot{p}} =
\frac{\dot{p}^2}{(C p+\dot{p})[C(1-p)-\dot{p}]} = K.
\label{eq:K}
\end{equation}
Here $K$ is the constant of integration. Solving the quadratic equation for
$\dot{p}$, we obtain two first order ODEs,
\begin{equation}
 \dot{p} = \frac{K(1-2p)\mp\sqrt{K^2+4Kp(1-p)}}{2C(1+K)} \label{eq:ode}.
\end{equation}
The upper sign ($-$) should be used for upward processes in which the quantum
level is raised and the lower sign ($+$) for downward processes. It is worth
mentioning a useful symmetry between electrons and holes. We are using the state
of an electron, $\epsilon(t)$ and $p(t)$, to describe the state of system.
Instead, we can also use the state of holes, $-\epsilon(t)$ and $1-p(t)$. If
$p(t)$ is a solution for an upward process, then $1-p(t)$ is a solution for a
downward process with $-\epsilon(t)$. Hence we do not need to calculate the
downward process separately, as it follows from this symmetry.

Before turning to the solution of the differential equation (\ref{eq:ode}), we
examine the physical meaning of the constant $K$. Eliminating $\dot{p}$ in
Eq.~(\ref{eq:K}) by using the master equation~(\ref{eq:master2}), the resulting
quadratic equation for $p(t)$ leads to the relation
\begin{equation}
 p(t) =
  \frac{1}{e^{\beta\epsilon(t)}+1}\left [1 + e^{\beta\epsilon(t)/2} \sqrt{K} \right ].
\label{eq:p}
\end{equation}
This equation indicates that when $K=0$, $p(t)$ is the equilibrium distribution
associated with the instantaneous value of the energy, implying that $K=0$
corresponds to the quasi-static limit ($\tau \rightarrow \infty$). As $K$
increases, $p(t)$ deviates from the equilibrium distribution.  We conclude that
$K$ measures how far the state of the system deviates from the quasi static
limit. We will use this insight below to obtain a perturbative solution for
small dissipation by assuming that $K$ is small.

Next we proceed to solve Eq.~(\ref{eq:ode}). Separation of the variables $p$ and
$t$ leads to the following explicit result for the upward processes:
\begin{equation}\label{eq:general}
C t= F[p(t);K]-F[p(0);K],
\end{equation}
where
\begin{equation}
 \begin{split}
 F(p;K) &= -\frac{1}{2}\ln p + \frac{1}{\sqrt{K}} \text{arctan} \left [ \frac{1-2p}{\sqrt{K+4p(1-p)}} \right ] \\
&+ \frac{1}{2} \ln \left [ \frac{2p+K+\sqrt{K^2 + 4Kp(1-p)}}{2(1-p)+K+\sqrt{K^2
+ 4Kp(1-p)}} \right ]. \label{eq:F}
\end{split}
\end{equation}
For the downward processes, we need to use $F(1-p;K)$.

The value of $K$ is determined by the boundary conditions:
\begin{equation}
 C \tau = F[p(\tau);K]-F[p(0);K].
\label{eq:boundary}
\end{equation}
Note that $K$ depends solely on the operation time $\tau$, the probabilities
$p_0$ and $p_1$, and the tunneling rate $C$ but not on temperature.
Unfortunately, the function (\ref{eq:F}) is quite complicated so we can not 
obtain an analytical expression for $K$. In general we need to solve for it
numerically. However, an exact perturbative solution is possible, cf. the next
section.

Having thus obtained the optimal $p(t)$ with $K$ determined by
(\ref{eq:general}), we insert this expression in Eq.~(\ref{eq:Q}) to obtain the
corresponding maximum heat for the optimal upward processes,
\begin{equation}
 \begin{split}
\beta \mathcal{Q} &= \int_0^\tau \epsilon(t) \dot{p} dt = \int_{p(0)}^{p(\tau)}
\epsilon(p) \md p \\
&= \int_{p(0)}^{p(\tau)} \md p \ln \left [
\frac{2p(1-p)+ K + \sqrt{K^2+4Kp(1-p)}}{2p^2} \right ]\\
& = \mathcal{S}[p(\tau);K]-\mathcal{S}[p(0);K] = \Delta\mathcal{S},
\label{eq:optQ}
\end{split}
\end{equation}
where
\begin{equation}\label{eq:S}
\begin{split}
 \mathcal{S}(p;K) &= p \ln \left [ \frac{2(1-p)p +K - \sqrt{K^2 + 4Kp(1-p)}}{2p^2} \right ]\\
 &- \sqrt{K} \arcsin \left [ \frac{1-2p}{\sqrt{K+1}} \right ] \\
&- \ln \left [\frac{2(1-p)-K-\sqrt{K^2 + 4Kp(1-p)}}{2} \right ] .
\end{split}
\end{equation}
For the downward processes, $\mathcal{S}(p;K)$ is replaced by
$\mathcal{S}(1-p;K)$.

Out of equilibrium, $\mathcal{S}$ is different from the system entropy
$S(p)=-p\ln p - (1-p)\ln (1-p)$. Indeed, $\Delta \mathcal{S}$ is the entropy
flow and is related to the system entropy change $\Delta S=S(p(\tau))-S(p(0))$
via the always-positive entropy production $\Delta_i S=\Delta S-\Delta
\mathcal{S} \geq 0$. It is only in the quasi-static limit, where $K\rightarrow
0$ and thus $\Delta_i S=0$, that $\mathcal{S}(p;K)$ reduces to $S(p)$.

We are now ready to apply the above results to our heat engine. To make the connection with the left/right symmetry required for the universality of the coefficient in the quadratic term, cf. the discussion in the introduction, it will be of interest to consider an asymmetry in the rate constant: we will use the subscripts  $C_c$ and $C_h$ for the  rate constant $C$  when in contact with the cold and hot reservoir, respectively. Recalling that processes I and III are upward and downward processes, respectively, the boundary condition (\ref{eq:boundary}) leads to
\begin{subequations}\label{eq:exactK}
\begin{eqnarray}
&& C_c \tau_c = F(p_1;K_c)-F(p_0;K_c) \\
&& C_h \tau_h = F(1-p_0;K_h)-F(1-p_1;K_h),
\end{eqnarray}
\end{subequations}
which determine the integration constants $K_c$ and $K_h$, respectively.

Substituting $K_c$ and $K_h$ into Eq. (\ref{eq:optQ}), we obtain the amount of heat that enters the system during the processes I and III:
\begin{subequations}\label{eq:exactQ}
 \begin{align}
 Q_c &= T_c [\mathcal{S}(p_1;K_c)-\mathcal{S}(p_0;K_c)] = T_c \Delta \mathcal{S}_c\\
 Q_h &= T_h [\mathcal{S}(1-p_0;K_h)-\mathcal{S}(1-p_1;K_h) = T_h \Delta
\mathcal{S}_h,
 \end{align}
\end{subequations}
which leads to the efficiency of the engine
\begin{equation}\label{eq:exact_eta}
 \eta = 1+ \frac{T_c \Delta \mathcal{S}_c}{T_h \Delta \mathcal{S}_h}.
\end{equation}
In the quasi-static limit, $K_c \rightarrow 0$ and $K_h \rightarrow 0$, one has
$\mathcal{S}(p;0)=S(p)=S(1-p)$, hence $\Delta \mathcal{S}_c=- \Delta
\mathcal{S}_h$, so that  Eq. (\ref{eq:exact_eta}) reduces to Carnot efficiency.

The above results provide the required optimization with respect to the
schedules. It remains to perform the optimization with respect to the remaining
degrees of freedom $\tau_c$, $\tau_h$, $p_0$ and $p_1$. In general, this can
only be done numerically since Eq. (\ref{eq:general}) only provides an implicit
equation for the time-dependence of the optimal schedule. We are, however,
mainly interested in the verification of universal features of efficiency at
maximum power. We therefore proceed with a perturbative analysis for which
analytic solutions can be obtained.

\section{Weak Dissipation Limit}

The deviation from Carnot efficiency can be investigated using the theory of linear irreversible thermodynamics where $T_h-T_c$ is assumed to be smaller than the  temperatures $T_h$ and $T_c$ of the reservoirs. However, for finite time thermodynamics, a different kind of expansion, directly related to the irreversibility caused by finite operation time, is more natural.  As mentioned in the previous section,  $K$ is a direct measure of the deviation from the quasi-static limit. Hence, it is natural to expand thermodynamic quantities in $K$. Since this is an expansion about the reversible case of zero dissipation, we will refer to this as the limit of weak dissipation.

We expand Eq. (\ref{eq:F}) in a series in $\sqrt{K}$. The leading
term is
\begin{equation}
   F(p;K)=\frac{\arcsin(1-2p)}{\sqrt{K}}\,.
\end{equation}
With this approximation, we are able to solve Eq.~(\ref{eq:exactK}) for $K$
to obtain
\begin{equation}\label{eq:K1}
 \sqrt{K_\alpha}=\frac{|\phi_1 - \phi_0|}{C_\alpha \tau_\alpha}, \quad (\alpha= c, h)
\end{equation}
where
\begin{equation}\label{eq:phi}
\phi_i=\arcsin(1-2p_i), \quad(i=0,1).
\end{equation}
 The present expansion is thus valid under the following condition of weak dissipation:
\begin{equation}
 C_\alpha \tau_\alpha \gg |\phi_1-\phi_0|, \quad (\alpha= c, h).
\label{eq:condition1}
\end{equation}
Note that it can easily be satisfied in our model since the right hand side is
bounded by $\pi$.

Once we find the value of $K$, the remaining calculation is straightforward.
Equation~(\ref{eq:general}) leads to the optimal protocols:
\begin{subequations}\label{eq:OptProtocol}
\begin{eqnarray}
 p_c(t)&=&\frac{1}{2}\left[1-\sin\left( \frac{t}{\tau_c}|\phi_1-\phi_0| - \phi_0\right)\right]\\
 p_h(t)&=&\frac{1}{2}\left[1+\sin\left( \frac{t}{\tau_h}|\phi_1-\phi_0| +
\phi_1\right)\right].
\end{eqnarray}
\end{subequations}

Expanding in a Taylor series with respect to $\sqrt{K}$, Eq. (\ref{eq:S}) is approximated by the two lowest order terms as
\begin{equation}\label{eq:approx_S}
 \mathcal{S}(p) = S(p) - \arcsin(1-2p)\sqrt{K}.
\end{equation}
Inserting the value of $K$, we obtain the maximum heat
\begin{subequations}\label{eq:OptQ}
\begin{eqnarray}
 Q_c^* &=& -T_c \Delta S -   T_c \frac{(\phi_1-\phi_0)^2}{C_c \tau_c} \\
 Q_h^* &=&  T_h \Delta S -   T_h \frac{(\phi_1-\phi_0)^2}{C_h \tau_h}
\end{eqnarray}
\end{subequations}
where $\Delta S = S(p_0)-S(p_1)$ is the reversible entropy change.  The second
term on the right hand side is the irreversible heat, which has to be small
under the condition (\ref{eq:condition1}) of weak dissipation.  In the
quasi-static limit ($\tau \rightarrow \infty$), the second term vanishes and the
efficiency (\ref{eq:exact_eta}) reaches the Carnot efficiency, as expected.

When the operation time is too short, the irreversible heat becomes dominant and
the net heat becomes negative. Equation (\ref{eq:OptQ}) indicates that positive
power can be obtained only if
\begin{equation}\label{eq:posP}
\frac{(T_h-T_c)\Delta S}{(\phi_1 - \phi_0)^2} > \frac{T_c}{C_c \tau_c}+\frac{T_h}{C_h \tau_h}\, .
\end{equation}
This inequality is consistent with the condition of the asymptotic expansion
(\ref{eq:condition1}) and can thus be satisfied even for a large temperature
difference.

So far, we have maximized the power only for the fixed operation times $\tau_c$
and $\tau_h$ and the boundary values $p_0$ and
$p_1$ of the occupation probabilities.
Now we further maximize the power
\begin{equation}\label{eq:P}
 P=\frac{Q_c+Q_h}{\tau_c+\tau_h}
\end{equation}
 with respect to the operation times.  It is easy to find that the power is a
maximum when
\begin{subequations}\label{eq:OptTau}
\begin{eqnarray}
\tau_c^* &=& \frac{2(\phi_1-\phi_0)^2 T_c (1+\sqrt{T_h C_c/T_c C_h})}{C_c \Delta S (T_h-T_c)}\\
\tau_h^* &=& \frac{2(\phi_1-\phi_0)^2 T_h (1+\sqrt{T_c C_h/T_h C_c})}{C_h \Delta S (T_h-T_c)}\, .
\end{eqnarray}
\end{subequations}
This optimization reflects the usual competition with the denominator of the power preferring faster operation whereas the numerator suggests a slower schedule to stay closer to Carnot efficiency.

For the asymptotic expansion to be valid, the optimal operation times must
satisfy the condition (\ref{eq:condition1}) of weak
dissipation. That is, for the process III the following inequality must be
satisfied:
\begin{equation}
 \frac{|\phi_1-\phi_0|}{C_h \tau_h^*} = \frac{\Delta S}{2 |\phi_1-\phi_0|}
\frac{T_h-T_c}{T_h(1+\sqrt{T_c C_h/T_h C_c)}} \ll 1.
\end{equation}
This can be achieved in two ways.  The first one corresponds to the usual condition for linear irreversible thermodynamics,  $(T_h-T_c)/T_h \ll 1$. The alternative  is $\left |\Delta S/ (\phi_1-\phi_0)\right | \ll 1$.    In this limit, our result remains valid even for large temperature differences.

With the optimized operation times (\ref{eq:OptTau}), the resulting power is written as a function of $p_0$ and $p_1$:
\begin{equation}\label{eq:Pmax}
 P=\frac{(T_h - T_c)^2}{4 (\sqrt{T_h/C_h}+\sqrt{T_c/C_c})^2} D(p_0,p_1),
\end{equation}
where
\begin{equation}\label{eq:D}
D(p_0,p_1)=\frac{\Delta S^2}{(\phi_1-\phi_0)^2}\, .
\end{equation}
The power reaches its maximum when $D(p_0,p_1)$ takes a maximum value,
$D_\text{max}=0.439$ at $p_0=p_1=0.0832$ or $p_0=p_1=0.9168$.    At these
conditions, the optimal operation time (\ref{eq:OptTau}) and the maximum heat
(\ref{eq:OptQ}) both vanish. However, the power remains finite.  This final
optimization thus leads to a singular and unrealistic situation.  We
note, however, that since Eq. (\ref{eq:D}) does not depend on the system
parameters, the efficiency does in fact not depend on this final optimization
step. Therefore, we proceed to evaluate efficiency without further reference to
optimal occupation probabilities.

Using the maximum heat (\ref{eq:OptQ}) and the optimal time (\ref{eq:OptTau}),
we finally obtain the following remarkable result for the efficiency at the
maximum power:
\begin{eqnarray}
 \eta^* &=& \frac{\eta_c(1+\sqrt{C_h T_c/C_c T_h})} {2(1+\sqrt{C_h
T_c/C_c T_h})-\eta_c},
\nonumber \\
&=&
\frac{\eta_c}{2}+\frac{\eta_c^2}{4(1+\sqrt{r})}+\frac{\eta_c^3}{8(1+\sqrt{r})}
+ o(\eta_c^4)
\label{eq:eta_maxP}
\end{eqnarray}
with $r=C_h/C_c$.
When $r=1$, the efficiency (\ref{eq:eta_maxP}) exactly coincides with the
Curzon-Ahlborn efficiency $\eta_{CA}=1-\sqrt{T_c/T_h}$.
Note also that the efficiency  is bounded below by $\eta_c/2$ for
$C_h /C_c \rightarrow \infty$ and bounded above by $\eta_c/(2-\eta_c)$ for
$C_h /C_c \rightarrow 0$. These limits can be realized without violating the
condition of weak dissipation.

\section{Discussion}

We have calculated the efficiency $\eta$ at maximum power of a Carnot cycle with
 a single level quantum dot as the operational device. Our calculation is in
agreement with known universality properties. In particular,  the efficiency
at maximum power is equal to half of the Carnot efficiency in the regime of
linear response, $\eta=\eta_{c}/2+...$. In the case of a left/right
symmetry, corresponding to equal exchange rate coefficients $C_h =C_c$ of the
dot with the heat reservoirs, the coefficient of the quadratic term is also
given by its universal value $1/8$, $\eta=\eta_{c}/2+\eta_{c}^2/8+...$.
However, we need to stress that this result was obtained not by an
expansion in $\eta_{c}$ but in the limit of weak dissipation. In fact, this
calculation adds a new perspective concerning the occurrence of Curzon-Ahlborn
efficiency itself. Indeed, in the presence of left/right symmetry, the
efficiency is actually {\it exactly} equal to the CA efficiency
$\eta_{CA}=1-\sqrt{1-\eta_{c}}$, in the limit of {\it weak dissipation}. This
limit is reminiscent of the original derivation of CA efficiency, and is in the
present model formally similar to the assumption of a linear conduction law
between reservoir and quantum dot. However the concept of weak dissipation is
more general. It remains to be explored whether this observation implies a wider
range of validity of  CA efficiency. In particular, it could explain why
observed efficiencies at maximum power are not very different from CA efficiency
in a wide range of systems under operational conditions far from linear
response.

\acknowledgments

M. E. is supported by the Belgian Federal Government (IAP project ``NOSY").
This research is supported in part by the NSF under grant PHY-0855471.


\end{document}